\documentclass[11pt,letter]{article}
\usepackage[top=1in,bottom=1in,left=1in,right=1in]{geometry}
\pdfoutput = 1
\usepackage[noEucal]{main}
\usepackage{aas_macros}
\usepackage[utf8]{inputenc}
\makeatletter \g@addto@macro\@floatboxreset\centering \makeatother
\usepackage[normalem]{ulem} 
\usepackage{graphicx}
\usepackage[scr=boondoxo]{mathalfa}
\usepackage[bbgreekl]{mathbbol}
\DeclareSymbolFontAlphabet{\mathbbm}{bbold}
\usepackage[font={small,it}]{caption}

\def\dS{{\mathsf{dS}}}

\def\U{{U}}
\def\SU{{SU}}

\def\CCFT{{\mathsf{CCFT}}}

\def\soft{{\mathsf{soft}}}
\def\hard{{\mathsf{hard}}}

\def\mat{{\mathsf{mat}}}

\def\gauge{{\mathsf{gauge}}}
\def\grav{{\mathsf{grav}}}

\def\eff{{\mathsf{eff}}}

\newcommand{\dbra}[1]{\langle\hspace{-0.11cm}\langle \,#1 \, |}
\newcommand{\dket}[1]{ | \, #1 \,\rangle\hspace{-0.11cm}\rangle}

\begin{document}
\begin{titlepage}
\unitlength = 1mm
~\\
\vskip 3cm
\begin{center}

{\LARGE{\textsc{A Holographic Model for Soft Photons and Gravitons in Four Dimensions}}}
\vspace{0.8cm}

Sangmin Choi${}^a$ and Prahar Mitra${}^{a,b}$

\vspace{1cm}
${}^a${\it Institute for Theoretical Physics, University of Amsterdam,
Science Park 904, Postbus 94485, 1090 GL Amsterdam, The Netherlands}\\
${}^b${\it School of Physics \& Astronomy, University of Southampton, Southampton SO17 1BJ, UK}

\vspace{0.8cm}

\begin{abstract}
We construct a two-dimensional action on the celestial sphere that describes the infrared sector of Abelian gauge and gravitational theories in four dimensions. In particular, we use the holographic model to reproduce (1) antipodal matching conditions for the superphaserotation and supertranslation Goldstone modes in four dimensions, (2) leading soft photon and graviton theorems, and (3) infrared factorization of amplitudes with generic dressed \emph{in} and \emph{out} states. Using (3), we reproduce the infrared divergences that plague the standard undressed amplitudes, and show that amplitudes involving Faddeev-Kulish dressed states are infrared finite. As a corollary, we use our holographic model to construct an infinite class of dressed states that give rise to infrared finite scattering amplitudes.

\end{abstract}

\vspace{1.0cm}
\end{center}
\end{titlepage}
\pagestyle{empty}
\pagestyle{plain}
\pagenumbering{arabic}

\tableofcontents

\section{Introduction}

The central conjecture of celestial holography \cite{Strominger:2017zoo, Pasterski:2021rjz, Raclariu:2021zjz, Prema:2021sjp} is that scattering amplitudes in four-dimensional asymptotically flat spacetimes are related, via an integral transform, to correlation functions of conformal primary operators in a two-dimensional CFT, known as the \emph{Celestial CFT} (CCFT),
\begin{equation}
\begin{split}
\label{integral_transform}
&\avg{ \CO_{h_1,{\bar h}_1}(x_1) \cdots \CO_{h_n,{\bar h}_n}(x_n) }_\CCFT \equiv \prod_{k=1}^n \left( \int_{\mrr^3} \dt^3 p_k \CK_{h_k,{\bar h}_k}{}^{\s_k} ( p_k | x_k) \right) A_{\s_1\cdots\s_n} ( p_1  , \cdots , p_n) .
\end{split}
\end{equation}
The LHS is a correlation function of conformal primary operators with weights $(h_k,{\bar h}_k)$, and the RHS is an $n$-point scattering amplitude involving particles with momenta $p^\mu_k$ ($p_k^2 = - m_k^2$) and little group spin index $\s_k$. The precise form of the kernel $\CK$ depends on mass and spin of the $k$th particle involved in the scattering process. The main issue with celestial holography, as it stands today, is that \eqref{integral_transform} is a definition, rather than an equivalence. In order to have a true holographic description of asymptotically flat spacetimes, one would like to obtain an independent definition of the CCFT that allows us, at least in principle, to evaluate the correlators on the LHS of \eqref{integral_transform} directly, i.e.\ without using the RHS. The conjecture may then be verified by showing that the two sides of \eqref{integral_transform}, independently evaluated, are equal.

In the absence of an independent definition of the CCFT, development in this field has been mostly bottom-up wherein one uses known properties of scattering amplitudes to determine properties of the CCFT, with the hope that once enough properties are determined, one might be able to ``guess'' the CCFT.\footnote{Top-down approaches towards flat holography has been developed in \cite{Costello:2020jbh, Costello:2021kiv, Costello:2022wso, Costello:2023hmi}. An alternative bottom-up approach to celestial holography that uses a different holographic dictionary has been developed in \cite{Sleight:2023ojm, Iacobacci:2024nhw}.} While this endeavour has not yet been fruitful to the extent hoped, we have made progress in constructing the ``universal'' part of the CCFT action. This universal piece, known as the \emph{soft effective action} (SEA), was constructed first in \cite{Kapec:2021eug} (see also \cite{Himwich:2020rro, Magnea:2021fvy, Nguyen:2021ydb, Gonzalez:2021dxw, Kalyanapuram:2021bvf}). The SEA is a two-dimensional action that lives on the celestial sphere $S^2$. It describes the dynamics of the soft and edge modes in gauge and gravitational theories. In \cite{Kapec:2021eug}, the SEA was constructed using ideas from effective field theory, namely by identifying the symmetry breaking patterns, the Goldstone and low energy modes, and then writing the most general consistent action order-by-order in a derivative expansion. A first principles derivation of this action was eventually provided for gauge theories in \cite{He:2024ddb} and (partially) for gravitational theories in \cite{He:2024vlp}. The SEA reproduces both soft theorems and infrared divergences in scattering amplitudes evaluated in a Fock basis of asymptotic states. It is well known, however, that in gauge and gravitational theories, the appropriate set of asymptotic states are Fock states dressed with clouds of soft photons or gravitons (a.k.a.\ the Faddeev-Kulish (FK) states \cite{Kulish:1970ut,Ware:2013zja}). The soft factorization properties of such amplitudes are also universal, but the SEA does not capture them. It also does not capture the antipodal matching conditions satisfied by the edge modes, which is another universal aspect of gauge and gravitational theories.

In this paper, we propose a generalization of the SEA that we refer to as the \emph{Generalized SEA} which reproduces all the features mentioned above. To reiterate, these are the (1) antipodal matching conditions, (2) infrared divergences in Fock states, (3) infrared divergences (or lack thereof) in FK dressed states (also for more general dressings), and (4) soft theorems. The generalized SEA (much like the SEA) is a Gaussian two-dimensional action, so correlators in such a theory are rather straightforward to evaluate. It is quite remarkable, however, that such elementary calculations in two-dimensions can reproduce rather complicated physics in the bulk four dimensions. For example, the derivation of infrared divergences requires the evaluation of a 1-loop Feynman integral (there are infinitely many diagrams that resum to the exponential of a 1-loop diagram \cite{Weinberg:1995mt}). The calculation that shows that scattering amplitudes in FK dressed states are infrared-finite is not trivial to say the least (e.g., see the twelve page long Appendix B of \cite{Choi:2017bna}); and the calculation that derives the matching condition requires blowing up spatial infinity into de Sitter spacetime and is similarly complicated (see \cite{Campiglia:2017mua}). The generalized SEA can reproduce these results in a page long calculation!

The paper is organized as follows. In Section \ref{sec:notations}, we setup the notation and conventions employed in this paper, and in Sections \ref{sec:soft_sector_gauge} and \ref{sec:soft_sector_grav}, we review the universal soft features of scattering amplitudes in gauge and gravitational theories, respectively. Detailed calculations relevant for these sections are relegated to Appendix \ref{app:cps_gauge}. In Section \ref{sec:gen_sea}, we introduce the generalized SEA and show that it reproduces all the soft features described in Section \ref{sec:prelims}. We summarize in Section \ref{sec:comments} and discuss future directions.

\section{Preliminaries}
\label{sec:prelims}

\subsection{Notation and Conventions}
\label{sec:notations}

To describe four-dimensional Minkowski spacetime $M_4$, we work in flat null coordinates $(u,x^a,r) \in \mrr^4$, which are related to the standard Cartesian coordinates $X^\mu=(t,\vec{x})$ via
\begin{equation}
\begin{split}
\label{flat_null_coord}
X^\mu = r {\hat q}^\mu(x) + 2 u n^\mu , \qquad {\hat q}^\mu (x) \equiv \left( \frac{1+x^a x_a}{2},x^a,\frac{1-x^a x_a}{2} \right) , \qquad n^\mu \equiv \left( \frac{1}{2},0^a,-\frac{1}{2} \right) .
\end{split}
\end{equation}
Lowercase Latin indices are valued in $\{1,2\}$ and are raised and lowered with the Euclidean metric $\d_{ab}$. In flat null coordinates, the metric of Minkowski spacetime is
\begin{equation}
\begin{split}
\label{Minkowski_metric}
\dt s^2 = - 2 \dt u \dt r + r^2 \dt x^a \dt x_a. 
\end{split}
\end{equation}
Future and past null infinities $\CI^\pm$ are located at $r \to \pm \infty$  with $(u,x^a)$ fixed. The past (future) boundary of $\CI^+$ ($\CI^-$) is at $u=-\infty$ ($u=+\infty$) and is denoted by $\CI^+_-$ ($\CI^-_+$). Importantly, the point labelled by $x^a$ on $\CI^+$ is antipodal to the point with the same label on $\CI^-$. These coordinates are explored further in Appendix A of \cite{He:2019jjk}.

An $n$-point scattering amplitude in $M_4$ is denoted by
\begin{equation}
\begin{split}
\label{scat_amp}
A_n = \avg{ \CO_1 \cdots \CO_n }_\mu , 
\end{split}
\end{equation}
where $\avg{\cdots} \equiv \bra{\O,+} \text{T} \{ \, \cdots \} \ket{\O,-}$ is the time-ordered vacuum correlation function and the \emph{plane-wave operator} $\CO_{\s_k}(p_k)$ is given by the LSZ reduction formula,
\begin{equation}
\begin{split}
\label{LSZ_def}
\CO_k \equiv \CO_{\s_k}(p_k) = - i u^A_{\s_k}(p_k) \int_{M_4} \dt^4 X e^{ - i p_k \cdot X } ( \p^2  - m_k^2 ) \varphi_A(X).
\end{split}
\end{equation}
The operator is labelled by an on-shell momentum $p_k^2 = - m_k^2$ and a little group spin index $\s_k$ ($\SU(2)$ if $m_k>0$ and $\U(1)$ if $m_k=0$). The spacetime field $\varphi_A(X)$ is labelled by a Lorentz spin index and a spacetime point $X^\mu$. The wavefunction $u^A_\s(p)$ describes the embedding of the plane wave operator into the spacetime field. In this paper, the momentum $p_k^\mu$ of a massive particle is parameterized as
\begin{equation}
\begin{split}
\label{mom_par}
p_k^\mu = m_k {\hat p}^\mu_k , \qquad {\hat p}^\mu_k \equiv z_k^{-1} {\hat q}^\mu(x_k) + z_k n^\mu . 
\end{split}
\end{equation}
Note that ${\hat p}_k^2 = - 1$. The operator $\CO_k(z_k,x_k)$ inserts an outgoing particle if $z_k > 0$ and an incoming particle if $z_k < 0$.

The subscript $\mu$ in \eqref{scat_amp} is the infrared cutoff that regulate the infrared divergences that plague four-dimensional Fock space scattering amplitudes. All photon or graviton loop integrals in Feynman diagrams are restricted to the region $|2 n \cdot \ell | > \mu$. This removes the $\ell^\mu=0$ region of the loop integrands which is where infrared divergences arise from.

\subsection{Soft Sector of Gauge Theories}
\label{sec:soft_sector_gauge}

In this section, we present a brief review of the soft sector of gauge theories. For completeness, any details omitted in the main text can be found in Appendix \ref{app:cps_gauge}.

\subsubsection{Vacuum Hilbert Space}
\label{sec:vac_HS_gauge}

We consider a $\U(1)$ gauge theory in $M_4$ that is described by a gauge field $A_\mu(X)$ and matter fields $\Phi_i(X)$ with charge $Q_i \in \mzz$. In the infrared, the theory is governed by Maxwell's equations
\begin{equation}
\begin{split}
\label{Maxwell_Eq}
\p^\mu F_{\mu\nu}(X) = e^2 J_\nu(X) , \qquad F_{\mu\nu}(X) = \p_\mu A_\nu(X) - \p_\nu A_\mu(X) ,
\end{split}
\end{equation}
where $e$ is the coupling constant of the gauge theory. The particle associated to the field $\Phi_i$ has electric charge $q_i = e Q_i$. The current $J_\mu(X)$ receives contributions from the charged matter fields $\Phi_i(X)$ as well as all possible higher derivative interaction terms in the Lagrangian. The theory is invariant under gauge transformations, which acts infinitesimally as
\begin{equation}
\begin{split}\label{gauge_transform}
\d_\l A_\mu(X) = \p_\mu \l(X) , \qquad \d_\l \Phi_i(X) = i Q_i \l (X) \Phi_i(X) , \qquad \l(X) \sim \l(X) + 2\pi . 
\end{split}
\end{equation}
For functions $\l$ that vanish at spatial infinity, \eqref{gauge_transform} represents a redundancy in our description of the theory. We remove these by working in axial null gauge
\begin{equation}
\begin{split}
\label{gauge_cond}
n^\mu A_\mu (X) = 0 .
\end{split}
\end{equation}
Gauge transformations that preserve \eqref{gauge_cond} and are non-vanishing on spatial infinity generate \emph{large gauge transformations} or \emph{superphaserotations}. These are physical symmetries of the theory and their Ward identities are the soft-photon theorems \cite{Weinberg:1965aa, Gell-Mann:1954wra, Low:1954kd, Low:1958sn, Burnett:1967km}.\footnote{Gauge transformations that are finite on $i^0$ are associated to the leading soft-photon theorem \cite{He:2014cra, Campiglia:2015qka, Kapec:2014zla, Kapec:2015ena, He:2019jjk} and those that diverge linearly are associated to the subleading soft-photon theorem \cite{Lysov:2014csa, Campiglia:2016hvg, Laddha:2017vfh, He:2019pll, He:2019ywq}.}

On a constant time slice, the gauge field admits a mode expansion of the form
\begin{equation}
\begin{split}
\label{mode_exp_gauge}
A_\mu(X) = e \int_{\mrr^3} \frac{\dt^3q}{(2\pi)^3} \frac{1}{2q^0} \left( a_\mu (t,q) e^{ i q \cdot X} + a_\mu^\dagger(t,q) e^{ - i q \cdot X} \right) , \quad n^\mu a_\mu(t,q) = 0 , \quad q^\mu = (|\vec{q}\,| , \vec{q}\,) . 
\end{split}
\end{equation}
At late times, 
\begin{equation}
\begin{split}
\label{a_large_t_gauge}
\lim_{t \to \pm \infty} a_\mu(t,q) = \ve_\mu^a(q) \CO_a^\pm(q) , \qquad n \cdot \ve_a(q) = q \cdot \ve_a(q) = 0 , \qquad \ve_a(q) \cdot \ve_b(q) = \d_{ab} .
\end{split}
\end{equation}
$\CO_a^\pm(q)$ is a canonically normalized \emph{in} ($-$) or \emph{out} ($+$) annihilation operator for the photon, so that
\begin{equation}
\begin{split}
\left[ \CO_a^\pm(q) , \CO_b^{\pm\dagger}(q') \right] = \d_{ab} ( 2\pi )^3 (2 q^0 ) \d^\3 ( q - q' ) . 
\end{split}
\end{equation}
We denote the Hilbert space of the theory on the Cauchy slices $\S^\pm = \CI^\pm \cup i^\pm$ by $\CH^\pm$. This factorizes into a vacuum or soft sector and a radiative or hard sector, $\CH^\pm \cong \CH_\soft^\pm \times \CH_\hard^\pm$. In this paper, we are interested in the vacuum sector of the gauge theory which is spanned by the \emph{soft-photon operator} $N_a^\pm$ and the \emph{superphaserotation Goldstone operator} $C_a^\pm$. These operators appear in the soft expansion of $\CO^\pm_a(q)$,
\begin{equation}
\begin{split}
\label{Oa_small_o}
\CO^\pm_a(\o {\hat q}(x)) = - \frac{4\pi}{e} \left( \frac{N^\pm_a(x)}{\o \pm i \e} - 2 \pi i C_a^\pm(x) \d(\o) \right) +  O(\ln (\o/\e) ) . 
\end{split}
\end{equation}
Under large gauge transformations \eqref{gauge_transform}, the soft operators transform as
\begin{equation}
\begin{split}\label{NC_LGT}
\d_{\ve^\pm} N_a^\pm(x) = 0 , \qquad \d_{\ve^\pm} C^\pm_a(x) = \p_a \ve^\pm(x) , \qquad \ve^\pm(x) \equiv \l|_{\p\S^\pm}(x) . 
\end{split}
\end{equation}
$N_a^\pm(x)$ and $C_a^\pm(x)$ are Hermitian and the commute with the Hamiltonian.
In this paper, we restrict our attention to gauge theories without magnetically charged states where\footnote{Magnetic charges are discussed in \cite{Strominger:2015bla, Choi:2019sjs, Kapec:2021eug, Mitra:2026xxx}.}
\begin{equation}
\begin{split}
\label{no_magnetic_charges}
N_a^\pm(x) = \p_a N^\pm(x) , \qquad C_a^\pm(x) = \p_a C^\pm(x) . 
\end{split}
\end{equation}
$N^\pm(x)$ and $C^\pm(x)$ are conjugate operators with (see Appendix \ref{app:soft_comm})
\begin{equation}
\begin{split}
\label{commutators_gauge}
[ C^\pm(x) , N^\pm(x') ] = - i e^2 \CG_\gauge(x-x') , \qquad \CG_\gauge(x-x') \equiv \frac{1}{4\pi} \ln[(x-x')^2] .
\end{split}
\end{equation}
Note that $\p^2 \CG_\gauge(x-x') = \d^\2(x-x')$. Composite operators that involve products of $N$ and $C$ are ordered by moving all the $N$'s to the right and all the $C$'s to the left.

To describe $\CH^\pm_\soft$, we work with eigenstates of $C^\pm(x)$, which satisfy
\begin{equation}
\begin{split}
\label{C_eigenstate_gauge}
C_a^\pm(x) \ket{\CC,\pm} &= \CC_a(x) \ket{\CC,\pm}  , \quad N_a^\pm(x) \ket{\CC,\pm} = - i e^2 \p_a \int_{\mrr^2} \dt^2 x' \CG_\gauge(x-x') \frac{\d}{\d \CC(x')} \ket{\CC,\pm} ,
\end{split}
\end{equation}
where $\CC_a(x) = \p_a \CC(x)$. More general vacuum states can be expanded in this basis and we denote the vacuum wave-function by $\Psi(\CC)$.

Eigenstates of $N^\pm(x)$ with eigenvalue $\CN(x)$ are given by
\begin{equation}
\begin{split}
\label{N_eigenstate}
\dket{\CN,\pm} = \int [ \dt \CC ] \exp \left( - \frac{i}{e^2} \int_{\mrr^2} \dt^2 x \, \CC^a(x) \CN_a (x) \right) \ket{\CC,\pm} . 
\end{split}
\end{equation}
It can be shown that the Lorentz-invariant vacuum state that is used in the definition of the scattering amplitude \eqref{scat_amp} is $\ket{\O,\pm} \equiv \dket{0,\pm}$ (see Appendix \ref{app:Lorentz_inv_vac_state}).

\subsubsection{Soft Factorization}
\label{sec:soft_factorization_gauge}

A scattering amplitude in a four-dimensional gauge theory factorizes into a universal soft factor and a non-universal hard factor that depends on the details of the theory,
\begin{equation}
\begin{split}
\label{soft_factorization_gauge}
A_{n+m} = \SS_m \wt{A}_n . 
\end{split}
\end{equation}
The soft factor $\SS_m$ receives contributions from external or ``real'' soft photons (governed by the leading soft-photon theorem) and internal or ``virtual'' soft photons (which give rise to infrared divergences). We discuss each of these contributions separately.

\paragraph{Leading Soft-Photon Theorem} The contribution of real soft photons is given by the leading soft-photon theorem \cite{Weinberg:1965aa},
\begin{equation}
\begin{split}
\label{soft_photon_thm}
\avg{ \CO_{a_1}(q_1) \cdots \CO_{a_m}(q_m) \CO_1 \cdots \CO_n }_\mu ~\xrightarrow{q_i^\mu \to 0}~ \prod_{i=1}^m \left( e \sum_{k=1}^n Q_k \frac{ p_k \cdot \ve_{a_i}(q_i)}{p_k \cdot q_i - i \e } \right) \avg{  \CO_1 \cdots \CO_n }_\mu .
\end{split}
\end{equation}
$\CO_a(q)$ is given via the LSZ reduction formula as
\begin{equation}
\begin{split}
\label{LSZ_def_gauge}
\CO_a(q) = - \frac{i}{e} \ve^\mu_a(q) \int_{M_4} \dt^4 X e^{ - i q \cdot X } \p^2 A_\mu (X).
\end{split}
\end{equation}
We parameterize the photon momentum as $q^\mu = \o {\hat q}^\mu(x)$. Then, in the gauge \eqref{gauge_cond}, the polarization vector is given by $\ve^\mu_a(q) = \p_a {\hat q}^\mu(x)$. It can then be shown that (see Appendix \ref{app:LSZ_bdy})
\begin{equation}
\begin{split}
\label{soft_limit_op_gauge}
\lim_{\o \to 0} [ \o \CO_a(\o {\hat q}(x)) ]  = - \frac{4\pi}{e} [ N_a^+(x)  - N_a^-(x) ] \equiv - \frac{4\pi}{e} N_a(x) . 
\end{split}
\end{equation}
Using the result above we can rewrite the soft theorem \eqref{soft_photon_thm} as
\begin{equation}
\begin{split}
\label{soft_thm_gauge}
\avg{ N_{a_1}(x_1) \cdots N_{a_m}(x_m) \CO_1 \cdots \CO_n }_\mu  = \CJ_{a_1}(x_1) \cdots \CJ_{a_m}(x_m) \avg{  \CO_1 \cdots \CO_n }_\mu ,
\end{split}
\end{equation}
where
\begin{equation}
\begin{split}
\label{Ja_def}
\CJ_a (x) \equiv \CJ_a^+(x)  + \CJ_a^-(x) , \qquad \CJ_a^\pm (x) \equiv - \frac{e^2}{4\pi} \p_a \sum_{k \in { \mathsf{out} \atop \mathsf{in} } } Q_k \ln ( {\hat p}_k \cdot {\hat q}(x) ) . 
\end{split}
\end{equation}
The soft theorem \eqref{soft_thm_gauge} was originally derived in \cite{Low:1958sn, Jauch:1976ava, Yennie:1961ad, Weinberg:1964ew} order-by-order in perturbation theory (i.e., by expanding the amplitude in powers of the coupling constant $e$). Recent work \cite{He:2014cra, Campiglia:2015qka, Kapec:2015ena} has shown that the soft theorem \eqref{soft_thm_gauge} is actually a consequence of the following two \emph{antipodal matching conditions}\footnote{The first antipodal matching condition is imposed by hand and is required to have a well-defined scattering amplitude. The second one is derived in Appendix \ref{app:matching_cond}. The extra minus sign in the second matching condition is due to our choice of coordinates (see Footnote 8 of \cite{He:2019jjk}).}
\begin{equation}
\begin{split}
\label{match_cond_gauge}
A_a |_{\p\S^+} (x) = A_a |_{\p\S^-} (x) , \qquad ( r^2 F_{ur} ) |_{\p\S^+}(x)  = - ( r^2 F_{ur} ) |_{\p\S^-}(x)  . 
\end{split}
\end{equation}
These are derived by blowing up spatial infinity $i^0$ into a de Sitter slice and then analyzing the structure of the Maxwell's equations on the slice \cite{Campiglia:2017mua, Esmaeili:2019hom}. The first matching condition breaks the large gauge symmetry \eqref{NC_LGT} down to a diagonal subgroup that is generated by the function $\ve(x) = \ve^+(x) = \ve^-(x)$. The Noether charge that generates this symmetry on $\S^\pm$ is given by
\begin{equation}
\begin{split}
\label{soft_charge_gauge}
Q^\pm_\ve = \pm \frac{1}{e^2} \oint_{\p\S^\pm} \dt^2 x \ve (r^2 F_{ur}) .
\end{split}
\end{equation}
The second matching condition in \eqref{match_cond_gauge} then implies that $Q_\ve^+ = Q_\ve^-$ which, as shown in \cite{He:2014cra}, is equivalent to the leading soft-photon theorem.

\begin{figure}[t!]
\begin{center}
\includegraphics[scale=0.35]{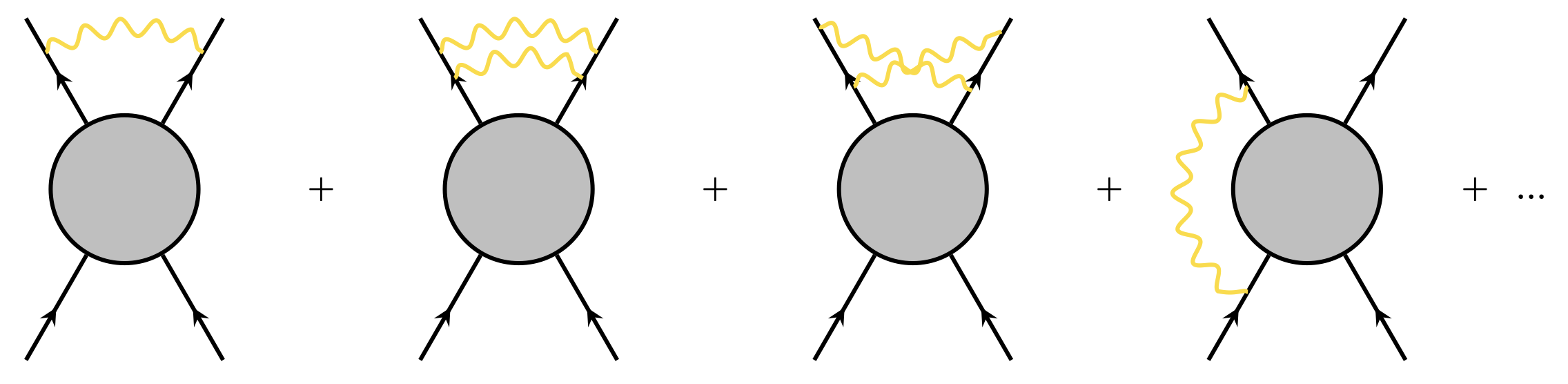}
\end{center}
\caption{Feynman Diagrams that contribute to Infrared Divergences \cite{Kapec:2021eug}}
\label{fig:Feynman_diagrams}
\end{figure}

\paragraph{Infrared Divergences} The contribution of virtual soft photons to the soft factor arises from diagrams of the form shown in Figure \ref{fig:Feynman_diagrams}.
Each diagram is individually divergent with the divergence arising from the $\ell^\mu=0$ region of the loop integral. The infinite sum of all the diagrams, however, vanishes, $A_n \sim e^{-\infty} = 0$. We regulate this divergence with the infrared cutoff $\mu$. It is a standard result in QFT that in the leading $\mu \to 0$ limit, the explicit sum of all the diagrams above takes the form (see Appendix \ref{app:cps_gauge})
\begin{equation}
\begin{split}
\label{IR_factorization_gauge}
A_n  = e^{-\G_\gauge} \wt{A}_n  ,  \qquad \G_\gauge = \frac{1}{2\pi e^2} \ln \frac{\L}{\mu} \int_{\mrr^2} \dt^2 x \CJ^a(x) \CJ_a(x) ,
\end{split}
\end{equation}
where $\wt{A}_n = \avg{ \CO_1 \cdots \CO_n }_\L$.\footnote{$\L$ is an intermediate scale that separates soft photons from hard photons. We assume that $\mu < \L \ll E_{\mathsf{typ}}$ where $E_{\mathsf{typ}}$ is the typical energy scale of the scattering amplitude. To remove the cutoff, we take $\mu \to 0$ followed by $\L \to 0$.} Combining this result with \eqref{soft_thm_gauge}, we find that the full soft factor $\SS_m$ is given by
\begin{equation}
\begin{split}
\label{soft_factor_gauge}
\SS_m = \CJ_{a_1}(x_1) \cdots \CJ_{a_m}(x_m)  \exp \left( -  \frac{1}{2\pi e^2} \ln \frac{\L}{\mu} \int_{\mrr^2} \dt^2 x \CJ^a(x) \CJ_a(x)  \right) . 
\end{split}
\end{equation}

\paragraph{FK Dressed States} It is clear from \eqref{soft_factor_gauge} that $\SS_m \to 0$ as $\mu \to 0$ so that the Fock space amplitude $A_{n+m}$ in \eqref{soft_factorization_gauge} vanishes (as expected)! The interpretation of this is rather simple -- the charged asymptotic states are \emph{not} Fock states in such theories. Rather, the correct asymptotic states are Fock states dressed with a cloud of soft photons. These dressed states are inserted by the operator $\wt{\CO}_k \equiv e^{R^\pm_k} \CO_k$ ($R_k^+$ for \emph{out} states and $R_k^-$ for \emph{in} states) where (see \cite{Gabai:2016kuf})
\begin{equation}
\begin{split}
\label{FK_dressing_k_gauge}
R^\pm_k &= e Q_k \int_{\mrr^3} \frac{\dt^3 q}{(2\pi)^3} \frac{1}{2q^0} f(q) \frac{p_k \cdot \ve^a (q)  }{ p_k \cdot q} \left( \CO^\pm_a(q) - \CO^{\pm\dagger}_a(q) \right)  , 
\end{split}
\end{equation}
where $f(q)$ is any function that satisfies $f(0) = \frac{1}{2}$ and is highly peaked near $q^\mu=0$ and falls of quickly away from this region.\footnote{The precise behaviour of $f(q)$ away from $q=0$ is not actually relevant as far as infrared finiteness of the corresponding amplitude is concerned. This only modifies the infrared finite part of the amplitude. In the language of edge and soft modes, changing $f(q)$ away from $q=0$ only modifies the dressing \eqref{FK_dressing_k_gauge_1} by hard operators, the contribution of which modifies the hard amplitude ${\wt A}_n$ in \eqref{IR_factorization_gauge}. } We take $f(0)=\frac{1}{2}$ (instead of $f(0)=1$) since photons of strictly zero energy have exactly one polarization, not two. An extended discussion of this factor of $1/2$ can be found around Eq. (40) of \cite{Gabai:2016kuf}. Such dressed states were studied first in non-relativistic theories by Dollard \cite{Dollard:1964cok} and later in QED by Kibble, Chung, Faddeev and Kulish \cite{Chung:1965zza, PhysRev.140.B1110, Kibble:1968sfb, Kibble:1968aa, Kibble:1968ab, Kibble:1968ac, Kulish:1970ut}, and are often referred to as FK states. It is rather non-trivial to show that scattering amplitudes evaluated in these dressed states are infrared finite, and this was shown in \cite{Gabai:2016kuf}.\footnote{There are, in fact, infinitely many dressed states that all lead to scattering amplitudes that are infrared finite. The dressing \eqref{FK_dressing_k_gauge} is the one proposed by Faddeev and Kulish in \cite{Kulish:1970ut}. We will discuss other possible dressings later in Section \ref{sec:gen_sea_gauge}.} Using \eqref{Oa_small_o}, it can be shown that \eqref{FK_dressing_k_gauge} simplifies to
\begin{equation}
\begin{split}
\label{FK_dressing_k_gauge_1}
R^\pm_k &=  \frac{i}{4 \pi} Q_k \int_{\mrr^2} \dt^2 x \left( C^{\pm a}(x) \pm \frac{1}{2} N^{\pm a}(x) \right)  \p_a \ln (  {\hat p}_k \cdot {\hat q}(x) ) . 
\end{split}
\end{equation}
Summing $R_k^+$ ($R_k^-$) over all \emph{out} (\emph{in}), and using the definitions \eqref{Ja_def}, we find
\begin{equation}
\begin{split}
\label{FK_dressing_gauge}
\sum_{k \in {\mathsf{out}\atop\mathsf{in}}} R^\pm_k &= - \frac{i}{e^2} \int_{\mrr^2} \dt^2 x \left( C^{\pm a}(x) \pm \frac{1}{2} N^{\pm a}(x) \right) \CJ^\pm_a(x) . 
\end{split}
\end{equation}

\subsection{Soft Sector of Gravitational Theories}
\label{sec:soft_sector_grav}

The soft sector of gravitational theories in four dimensions is qualitatively identical to that of gauge theories discussed in the previous section (modulo some factors of 2 and an extra tensor index). For this reason, we simply present the formulas that will be relevant for this paper. 

\subsubsection{Vacuum Hilbert Space}
\label{sec:vac_HS_grav}

Consider a gravitational theory in $M_4$ that is described by the graviton field $h_{\mu\nu}(X)$ and matter fields $\Phi_i(X)$. We suppress all the spin indices on the matter fields for clarity. In the infrared, the theory is governed by Einstein's equations
\begin{equation}
\begin{split}
R_{\mu\nu} - \frac{1}{2} g_{\mu\nu} R = \frac{1}{4} \k^2 T_{\mu\nu} , \qquad \k = \sqrt{32\pi G_N} ,
\end{split}
\end{equation}
where $R_{\mu\nu}$ is the Ricci tensor associated to the metric $g_{\mu\nu} = \eta_{\mu\nu} + \k h_{\mu\nu}$. The stress tensor receives contributions from all the matter fields as well as any higher derivative interactions in the Lagrangian. The theory is invariant under diffeomorphisms, which acts infinitesimally as
\begin{equation}
\begin{split}
\label{diff_transform}
\d_\xi h_{\mu\nu}(X) = \CL_\xi h_{\mu\nu}(X) + \p_\mu \xi_\nu(X) + \p_\nu \xi_\mu(X) , \qquad \d_\xi \Phi_i(X) =  \CL_\xi \Phi_i(X) . 
\end{split}
\end{equation}
Vector fields that vanish at $i^0$ correspond to redundancies in our description of the system. We remove these by working in flat Bondi-Sachs gauge where
\begin{equation}
\begin{split}
\label{Bondi_gauge_cond}
h_{rr} = h_{ra} = 0 , \qquad \det ( r^2 \d_{ab} + \k h_{ab} ) = r^4.
\end{split}
\end{equation}
Gauge transformations that preserve \eqref{Bondi_gauge_cond} and are non-vanishing on $i^0$ generate large diffeomorphisms. There are two classes of such diffeomorphisms: (1) \emph{Supertranslations} are parameterized by a function $f^\pm(x)$ with $\xi |_{\CI^\pm} = f^\pm \p_u$ and (2) \emph{Superrotations} are parameterized by a conformal Killing vector $Y^{\pm a}(x)$ with $\xi |_{\CI^\pm} = \frac{1}{2} \p \cdot Y^\pm ( u \p_u  - r \p_r ) + Y^{\pm a} \p_a$. These are physical symmetries of the theory and their Ward identities are the leading and subleading soft-graviton theorems, respectively \cite{Weinberg:1965aa, Cachazo:2014fwa, Laddha:2017ygw, Campiglia:2016efb, Campiglia:2016jdj}.\footnote{At loop level, superrotation Ward identities are the logarithmic soft graviton theorems \cite{Sahoo:2018lxl,Donnay:2022hkf,Agrawal:2023zea,Choi:2024ygx,Choi:2024ajz}.}

On a constant time slice, the graviton field admits a mode expansion of the form
\begin{equation}
\begin{split}
\label{mode_exp_grav}
h_{\mu\nu}(X) = \int_{\mrr^3} \frac{\dt^3q}{(2\pi)^3} \frac{1}{2q^0} \left( h_{\mu\nu} (t,q) e^{ i q \cdot X} + h_{\mu\nu}^\dagger(t,q) e^{ - i q \cdot X} \right) , \quad \lim_{t \to \pm \infty} h_{\mu\nu}(t,q) = \ve^{ab}_{\mu\nu}(q) \CO^\pm_{ab}(q) .
\end{split}
\end{equation}
$\CO_{ab}^\pm(q)$ is a canonically normalized \emph{in} ($-$) or \emph{out} ($+$) annihilation operator for the graviton.

As in gauge theories, the Hilbert space $\CH^\pm$ on $\S^\pm$ factorizes into a soft and a hard sector. The soft or vacuum sector of the theory is spanned by the \emph{soft-graviton operator} $N_{ab}^\pm$ and the \emph{supertranslation Goldstone operator} $C_{ab}^\pm$ which appear in the soft expansion of $\CO_{ab}^\pm(q)$,
\begin{equation}
\begin{split}
\label{Oab_small_o}
\CO^\pm_{ab}(\o {\hat q}(x)) = - \frac{4\pi}{\k} \left( \frac{N^\pm_{ab}(x)}{\o\pm i \e} - 2\pi i C_{ab}^\pm(x) \d(\o) \right) + O(\ln\o) .
\end{split}
\end{equation}
Under supertranslations \eqref{diff_transform},
\begin{equation}
\begin{split}
\label{supertranslations}
\d_{f^\pm} N^\pm_{ab}(x) = 0 , \qquad \d_{f^\pm} C^\pm_{ab}(x) = - 2 \p_{\{a} \p_{b\}} f^\pm(x).
\end{split}
\end{equation}
$N_{ab}^\pm$ and $C_{ab}^\pm$ are Hermitian operators that commute with the Hamiltonian and take the form
\begin{equation}
\begin{split}
N^\pm_{ab}(x) = - 2 \p_{\{a} \p_{b\}} N^\pm(x) , \qquad C^\pm_{ab}(x) = - 2 \p_{\{a} \p_{b\}} C^\pm(x) . 
\end{split}
\end{equation}
$C^\pm(x)$ and $N^\pm(x)$ are conjugate operators with
\begin{equation}
\begin{split}
\label{commutators_grav}
\left[ C^\pm(x) , N^\pm(x') \right] &= \frac{i\k^2}{2} \CG_\grav(x-x')  ,  \qquad \CG_\grav(x-x') \equiv \frac{1}{16\pi} (x-x')^2 \ln [ (x-x')^2 ].
\end{split}
\end{equation}
Note that $ (\p^2)^2 \CG_\grav(x-x') = \d^\2(x-x')$. To describe the soft Hilbert space, we work with eigenstates of $C^\pm(x)$, which satisfies
\begin{equation}
\begin{split}
C^\pm(x) \ket{\CC,\pm} &= \CC(x) \ket{\CC,\pm} , \qquad N^\pm(x) \ket{\CC,\pm} = \frac{i \k^2}{2} \int_{\mrr^2} \dt^2 x' \CG_\grav(x-x') \frac{\d}{\d \CC(x')}  \ket{\CC,\pm}  .
\end{split}
\end{equation}
Eigenstates of $N^\pm(x)$ with eigenvalue $\CN(x)$ are given by
\begin{equation}
\begin{split}
\dket{\CN,\pm} = \int[\dt \CC] \exp \left( \frac{i}{\k^2} \int_{\mrr^2} \dt^2 x \, \CC^{ab}(x) \CN_{ab} (x) \right) \ket{\CC,\pm} .
\end{split}
\end{equation}
As in gauge theory, the Lorentz invariant vacuum state used in the definition of scattering amplitude \eqref{scat_amp} is $\ket{\O,\pm} \equiv \dket{0,\pm}$.

\subsubsection{Soft Factorization}
\label{sec:soft_factorization_grav}

Scattering amplitudes in gravitational theories admit a universal factorization just like in gauge theories \eqref{soft_factorization_gauge}, and the soft factor $\SS_m$ receives contributions from both real and virtual soft gravitons.

\paragraph{Leading Soft-Graviton Theorem} The contribution of real soft-gravitons is given by the leading soft-graviton theorem \cite{Weinberg:1965aa}
\begin{equation}
\begin{split}
\label{soft_grav_thm}
\avg{ \CO_{a_1b_1}(q_1) \cdots \CO_{a_mb_m}(q_m) \CO_1 \cdots \CO_n }_\mu ~ \xrightarrow{q_i^\mu \to 0} ~ \prod_{i=1}^m \left(  \frac{\k}{2} \sum_{k=1}^n \frac{ p_{k\mu} p_{k\nu} \ve^{\mu\nu}_{a_ib_i}(q_i)}{p_k \cdot q_i - i \e } \right) \avg{  \CO_1 \cdots \CO_n }_\mu .
\end{split}
\end{equation}
Following the same procedure as in gauge theory, it can be shown that we can rewrite this as
\begin{equation}
\begin{split}
\label{soft_thm_grav}
 \avg{ N_{a_1b_1}(x_1) \cdots N_{a_mb_m}(x_m)\CO_1 \cdots \CO_n }_\mu = \CJ_{a_1b_1}(x_1) \cdots \CJ_{a_mb_m}(x_m) \avg{  \CO_1 \cdots \CO_n }_\mu ,
\end{split}
\end{equation}
where $N_{ab}(x) \equiv N_{ab}^+(x) - N_{ab}^-(x)$, and
\begin{equation}
\begin{split}
\label{Jab_def}
\CJ_{ab}(x) = \CJ_{ab}^+(x) + \CJ^-_{ab}(x) , \qquad \CJ_{ab}^\pm(x) \equiv - \frac{\k^2}{8\pi} \p_{\{a} \p_{b\}}  \sum_{k \in { \mathsf{out} \atop \mathsf{in} } } m_k ( {\hat p}_k \cdot {\hat q}(x) ) \ln ( {\hat p}_k \cdot {\hat q}(x) ) . 
\end{split}
\end{equation}
It was recently shown that this soft theorem \eqref{soft_thm_grav} is a consequence of the antipodal matching conditions similar to \eqref{match_cond_gauge},
\begin{equation}
\begin{split}
\label{match_cond_grav}
( r^{-1} h_{ab} )|_{\p\S^+} (x) = ( r^{-1} h_{ab} )|_{\p\S^-}  (x) , \qquad ( r^3 W_{urur} ) |_{\p\S^+}(x)  =  ( r^3 W_{urur} ) |_{\p\S^-}(x)  ,
\end{split}
\end{equation}
where $W_{\mu\nu\rho\s}$ is the Weyl tensor associated to the metric $g_{\mu\nu} =\eta_{\mu\nu}+\k h_{\mu\nu}$. The first matching condition in \eqref{match_cond_grav} breaks the supertranslation symmetry \eqref{supertranslations} down to a diagonal one generated by $f(x) = f^+(x) = f^-(x)$. The Noether charge that generates this symmetry on $\S^\pm$ is
\begin{equation}
\begin{split}
T_f^\pm = - \frac{4}{\k^2} \oint_{\p\S^\pm} \dt^2 x f (r^3 W_{urur}) . 
\end{split}
\end{equation}
The second matching condition in \eqref{match_cond_grav} then implies that $T_f^+ = T_f^-$ which, as shown in \cite{He:2014laa}, is equivalent to the leading soft-graviton theorem.

\paragraph{Infrared Divergences and FK States} Just as in gauge theories, the Fock space scattering amplitude in gravitational theories is infrared-divergent due to the same class of diagrams as Figure \ref{fig:Feynman_diagrams}. In the leading $\mu \to 0$ limit, the amplitude factorizes as \eqref{IR_factorization_gauge}, with
\begin{equation}
\begin{split}
\label{Gamma_grav}
A_n  = e^{-\G_\grav} \wt{A}_n , \qquad \G_\grav &= \frac{1}{2\pi\k^2} \ln \frac{\L}{\mu} \int_{\mrr^2} \dt^2 x \CJ^{ab}(x) \CJ_{ab}(x)  . 
\end{split}
\end{equation}
Combining this result with \eqref{soft_thm_grav}, we find that the full soft factor is 
\begin{equation}
\begin{split}
\label{IR_factor_grav}
\SS_m = \CJ_{a_1b_1}(x_1) \cdots \CJ_{a_mb_m}(x_m) \exp \left( - \frac{1}{2\pi\k^2} \ln \frac{\L}{\mu}  \int_{\mrr^2} \dt^2 x \CJ^{ab}(x) \CJ_{ab}(x) \right) . 
\end{split}
\end{equation}
This vanishes as $\mu \to 0$. The interpretation here is the same as in gauge theories -- the appropriate asymptotic states are Fock states that are dressed with a cloud of soft gravitons. The dressed operator is $e^{R_k^\pm} \CO_k$ where \cite{Ware:2013zja, Choi:2017bna, Choi:2017ylo}
\begin{equation}
\begin{split}
\label{FK_dressing_k_grav}
R^\pm_k &= \frac{\k}{2} \int_{\mrr^3} \frac{\dt^3 q}{(2\pi)^3} \frac{1}{2q^0} f(q) \frac{p_k^\mu p_k^\nu \ve_{\mu\nu}^{ab} (q)  }{ p_k \cdot q} \left( \CO^\pm_{ab}(q) - \CO^{\pm\dagger}_{ab}(q) \right)  .
\end{split}
\end{equation}
Using \eqref{Oab_small_o} and summing over \emph{in} or \emph{out} states, we find
\begin{equation}
\begin{split}
\label{FK_dressing_grav}
\sum_{k \in {\mathsf{out} \atop \mathsf{in}}} R^\pm_k &=  - \frac{i}{\k^2}  \int_{\mrr^2} \dt^2 x \left( C^{\pm ab}(x) \pm \frac{1}{2} N^{\pm ab}(x) \right)  \CJ^\pm_{ab}(x) .
\end{split}
\end{equation}

\section{The Generalized Soft Effective Action}
\label{sec:gen_sea}

In the previous section, we reviewed a few universal features of scattering amplitudes in four-dimensional gauge and gravitational theories. In particular, we highlight the following properties:
\begin{itemize}
    \item[(1)] Matching conditions: These are given in \eqref{match_cond_gauge} and \eqref{match_cond_grav}. The first of these matching conditions is equivalent to $C_a^+(x) = C_a^-(x)$ in gauge theories and $C_{ab}^+(x) = C_{ab}^-(x)$ in gravity.

    \item[(2)] Leading soft theorems: These are given in \eqref{soft_thm_gauge} and \eqref{soft_thm_grav}. These are equivalent to the second set of matching conditions in \eqref{match_cond_gauge} and \eqref{match_cond_grav}.

    \item[(3)] Infrared Divergences in undressed amplitudes: These are given in \eqref{IR_factorization_gauge} and \eqref{Gamma_grav}. It follows that scattering amplitudes evaluated in Fock states are zero.

    \item[(4)] FK Dressed dressed States: These are described in \eqref{FK_dressing_gauge} and \eqref{FK_dressing_grav}. Scattering amplitudes evaluated in these states are infrared-finite and non-vanishing. More generally, there is an infinite-family of dressed states that give rise to infrared-finite amplitudes.
    \end{itemize}
These properties are rather tricky to derive in four dimensions and are in no way ``obvious'' features of scattering amplitudes. In this section, we provide a holographic interpretation of these properties by constructing a two-dimensional action on the celestial sphere (a.k.a. the generalized SEA) that reproduces the properties (1)--(4). The SEA constructed in \cite{Kapec:2021eug} is a two-dimensional theory on the celestial sphere that governs the dynamics of the soft ($N^\pm$) and edge ($C^\pm$) modes in gauge and gravitational theories. It can be obtained from taking the soft limit of the on-shell bulk action, and the one that governs the leading soft modes has been derived in \cite{He:2024ddb, He:2024vlp}. We follow the derivation presented in \cite{He:2024ddb} to obtain the generalized SEA. Amazingly, it will turn out that the generalized SEA is Gaussian, so the holographic derivation of (1)--(4) is remarkably compact.

\subsection{Gauge Theory}
\label{sec:gen_sea_gauge}

We propose the following generalization of the SEA in Abelian gauge theories,\footnote{In this paper, we define a Lorentzian action so our path integral measure is $e^{ i S_\eff } $. On the other hand, previous literature on this topic \cite{Kapec:2021eug, He:2024skc} defined a Euclidean action and their path integral measure was $e^{ - S_{\mathsf{KM}}}$. The relationship between the two actions is $S_\mathsf{KM} = - i S_\eff$.}
\begin{equation}
\begin{split}
\label{gen_sea_gauge}
\!\!\! S_\eff[\CC^\pm,\CN^\pm] &= \frac{i}{2\pi e^2}\ln\frac{\L}{\mu} \int_{\mrr^2} \dt^2x \,\CN_a(x)\CN^a(x) - \frac{1}{e^2}\int_{\mrr^2} \dt^2 x \left( \CC^{+a}(x)\CN^+_a(x)- \CC^{-a}(x)\CN^-_a(x) \right),
\end{split}
\end{equation}
where
\begin{equation}
\label{soft_mode_diff}
    \CN_a(x) = \CN_a^+(x) - \CN_a^-(x).
\end{equation}
Note that here we use the calligraphic font $\CC^\pm$ and $\CN^\pm$ as the path integral is over all possible \emph{eigenvalues} of the operators $C^\pm$ and $N^\pm$. We will see the importance of this in Section \ref{sec:sea_fk}.

\eqref{gen_sea_gauge} can be derived be evaluating the contribution of the edge and soft modes to the Maxwell action,
\begin{equation}
    \begin{split}
    \label{Maxwell_action}
        S_\gauge[A] = - \frac{1}{2e^2} \int_{M_4} F \w \star F + \frac{1}{e^2} \int_{\CI^+} A \w \star F - \frac{1}{e^2} \int_{\CI^-} A \w \star F . 
    \end{split}
\end{equation}
The boundary terms above are required to have a well-defined variational principle with our boundary conditions (radiative boundary conditions on $\CI$ and Dirichlet boundary conditions on $i^0$). The precise details of this procedure is outlined in \cite{He:2024ddb} so we do not reproduce it here. The first, second, and third terms in \eqref{Maxwell_action} reproduce the first, second, and third terms in \eqref{gen_sea_gauge}, respectively.\footnote{The derivations in \cite{He:2024ddb} are given using a non-zero current. We expect a similar derivation to hold for the photon sector without any current.} The generalization employed in this paper lies in making the distinction between the future and past Goldstone modes $\CC_a^+$ and $\CC_a^-(x)$ respectively. The authors of \cite{He:2024ddb} put in the matching condition $\CC_a^+(x) = \CC_a^-(x)$ by hand when deriving the SEA.

The SEA is constructed to reproduce the universal soft factor $\SS_m$ \eqref{soft_factorization_gauge} in the scattering amplitude $A_{n+m}$. The precise form of this soft factor depends on the \emph{in} and \emph{out} vacuum state that is used to evaluate the amplitudes $A_{n+m}$. More precisely the statement is
\begin{equation}
\begin{split}
\label{soft_pi}
\bra{\CC^+,+} N_{a_1} \cdots N_{a_m}  \CO_1\cdots\CO_n \ket{\CC^-,-} |_{\soft} = \int [\dt \CN^+][\dt \CN^-]
e^{iS_\eff[\CC^\pm,\CN^\pm]} \CN_{a_1} \cdots \CN_{a_m} \, \CU_1 \cdots \CU_n .
\end{split}
\end{equation}
On the LHS, we have the soft part $\SS_m$ of a general scattering amplitude evaluated in eigenstates of $C^\pm$. We insert $m$ soft operators $N_{a_i}^\pm$ and $n$ of hard operators $\CO_k$. The universal soft part of this amplitude is given by the path integral shown above. For each hard operator $\CO_k$, we insert the Wilson line operator $\CU_k=\exp \big( iQ_k \int_{\g_k} A \big)|_\soft$, where $\g_k$ is the worldline of the $k$th hard particle. It was shown in \cite{Kapec:2021eug, He:2024skc} that when summing over all hard particles, this reduces to
\begin{align}
\CU_1 \cdots \CU_n &= \exp\left( \frac{i}{e^2} \int_{\mrr^2} \dt^2x [ \CC^{+a}(x)\CJ^+_a(x) + \CC^{-a}(x)\CJ^-_a(x) ]  \right)  .
\end{align}
Generalizing \eqref{soft_pi} to arbitrary \emph{in} and \emph{out} vacuum states, we find
\begin{equation}
\begin{split}
\label{soft_pi_gen}
&\bra{\Psi_\mathsf{out},+} N_{a_1} \cdots N_{a_m} \CO_1\cdots\CO_n \ket{\Psi_\mathsf{in},-} |_\soft \\
&\qquad \qquad = \int [\dt \CC^+][\dt \CC^-][\dt \CN^+][\dt \CN^-] e^{iS_\eff[\CC^\pm,\CN^\pm]} \Psi_\mathsf{out}^*(\CC^+) \Psi_\mathsf{in}(\CC^-) \CN_{a_1} \cdots \CN_{a_m} \, \CU_1 \cdots \CU_n ,
\end{split}
\end{equation}
where $\Psi_\mathsf{in}$ and $\Psi_\mathsf{out}$ are the wavefunctions associated with the \emph{in}- and \emph{out}-vacua respectively.

In the rest of this section, we show that this path integral correctly reproduces (1)--(4).

\subsubsection{Antipodal Matching Condition}
\label{sec:sea_amc}

We start with the antipodal matching condition, namely
\begin{equation}
    \begin{split}
    \label{match_cond_gauge_1}
        C^+_a(x) = C_a^-(x).
    \end{split}
\end{equation}
To derive this, we define
\begin{equation}
    \begin{split}
    \label{soft_mode_avg}
        \CN_a^\mathsf{avg}(x) \equiv \frac{1}{2} [ \CN_a^+(x) + \CN_a^-(x) ] .
    \end{split}
\end{equation}
In terms of the difference mode $\CN_a(x)$ \eqref{soft_mode_diff} and the average mode, the generalized SEA \eqref{gen_sea_gauge} takes the form
\begin{equation}
    \begin{split}
        \label{gen_sea_gauge_1}
    S_\eff[\CC^\pm,\CN,\CN^\mathsf{avg}]
    &= \frac{i}{2\pi e^2}\ln\frac{\L}{\mu}
		\int_{\mrr^2} \dt^2x \,\CN_a(x)\CN^a(x)
         - \frac{1}{2e^2}\int_{\mrr^2} \dt^2 x \left(  \CC^{+a}(x) 
			 +  \CC^{-a}(x) \right) \CN_a(x)  \\
         &\qquad \qquad - \frac{1}{e^2}\int_{\mrr^2} \dt^2 x \left( \CC^{+a}(x) -  \CC^{-a}(x) \right) \, \CN^\mathsf{avg}_a(x) .
    \end{split}
\end{equation}
We see that $\CN_a^\mathsf{avg}(x)$ appears linearly in the generalized SEA, which implies that it is non-dynamical. Moreover, scattering amplitudes are obtained from the correlation function through the LSZ reduction formula, which only allows insertions of the difference $\CN_a=\CN_a^+-\CN_a^-$ (see \eqref{soft_limit_op_gauge}). It follows that, as far as scattering processes are concerned, we may integrate out $\CN_a^\mathsf{avg}$, which then yields a Dirac delta functional $\d ( \CC^+_a -  \CC^-_a )$ which precisely implements the matching condition \eqref{match_cond_gauge_1}.

\subsubsection{Soft Theorem}
\label{sea_sth}

We next turn to soft theorem \eqref{soft_thm_gauge} which we can equivalently write as
\begin{equation}
    \begin{split}
    \label{soft_thm_gauge_1}
        \CN_a(x) = \CJ_a(x) = \CJ_a^+(x) + \CJ^-_a(x). 
    \end{split}
\end{equation}
To show this, we note that the leading soft theorem is derived in the Lorentz-invariant vacuum state that is assumed in standard QFT. These are eigenstates of $N^\pm$ with value $\CN_a(x) = 0$. From \eqref{N_eigenstate}, we see that the wave-function of these states is $\Psi(\CC) = 1$. The path integral \eqref{soft_pi_gen} then simplifies to (after integrating out $\CN^\mathsf{avg}$ as discussed in the previous section)
\begin{equation}
\begin{split}
\label{IR_pi_1}
&\dbra{0,+} N_{a_1} \cdots N_{a_m} \CO_1\cdots\CO_n \dket{0,-} |_\soft \\
&\qquad \qquad = \int [\dt \CC][\dt \CN] \exp \left(  - \frac{1}{2\pi e^2}\ln\frac{\L}{\mu}
\int_{\mrr^2} \dt^2x \,\CN_a(x)\CN^a(x) \right) \\
&\qquad \qquad \qquad \qquad \qquad \times \exp \left( - \frac{i}{e^2}\int_{\mrr^2} \dt^2 x \CC^\mathsf{avg}_a(x) [ \CN^a(x) - \CJ^a(x) ] \right) \CN_{a_1} \cdots \CN_{a_m} ,
\end{split}
\end{equation}
where $\CC^\mathsf{avg}_a \equiv \frac{1}{2} ( \CC^+_a + \CC^-_a )$ is the averaged edge mode. We see that $\CC_a(x)$ appears linearly in the path integral. Integrating this out yields a Dirac delta functional $\d(\CN_a-\CJ_a)$ which is precisely the soft theorem \eqref{soft_thm_gauge_1}.

\subsubsection{Infrared Divergences}
\label{sea_id}

The infrared divergent factorization of scattering amplitudes follows immediately from \eqref{IR_pi_1}. The path integral over $\CC$ produces the Dirac delta functional $\d(\CN_a-\CJ_a)$, which allows us to perform the path integral over $\CN$ trivially and we find 
\begin{equation}
\begin{split}
\label{IR_pi_2}
\!\!\! \dbra{0,+} N_{a_1} \cdots N_{a_m} \CO_1\cdots\CO_n \dket{0,-} |_\soft = \exp \left(  - \frac{1}{2\pi e^2}\ln\frac{\L}{\mu}
\int_{\mrr^2} \dt^2x \,\CJ_a(x)\CJ^a(x)
 \right)  \CJ_{a_1} \cdots \CJ_{a_m} . 
\end{split}
\end{equation}
The factors of $\CJ_a$ are associated to the soft theorem. The infrared divergent factor in the exponential is precisely the one in \eqref{IR_factorization_gauge}.

\subsubsection{Faddeev-Kulish Dressed States}
\label{sec:sea_fk}

As discussed in Section \ref{sec:soft_factorization_gauge}, scattering amplitudes in the FK dressed states are obtained by inserting the operators $\wt{\CO}_k \equiv e^{R_k^\pm} \CO_k$ in the amplitude (instead of $\CO_k$) where $R_k^\pm$ is defined in \eqref{FK_dressing_k_gauge_1}. This modifies the path integral \eqref{soft_pi_gen} by replacing the operators $\CU_k$ with $e^{R_k^\pm} \CU_k$, so the universal soft part of a dressed amplitude is given by
\begin{equation}
\begin{split}
\label{soft_pi_gen_FK}
\dbra{0,+} \wt{\CO}_1\cdots \wt{\CO}_n \dket{0,-} |_\soft = \int [\dt \CC^+][\dt \CC^-][\dt \CN^+][\dt \CN^-] e^{iS_\eff[\CC^\pm,\CN^\pm]} 
e^{\CR^{\eta_1}_1 + \cdots + \CR^{\eta_k}_n} \CU_1 \cdots \CU_n ,
\end{split}
\end{equation}
where $\eta_k = +1$ ($\eta_k = -1$) if the $k$th particle is outgoing (incoming). Here, we have taken the \emph{in} and \emph{out} vacuum to be the Lorentz-invariant vacuum states for simplicity. Using the explicit form of $R_k^\pm$ in \eqref{FK_dressing_k_gauge_1}, this simplifies to
\begin{equation}
\begin{split}
& \dbra{0,+} \wt{\CO}_1\cdots \wt{\CO}_n \dket{0,-} |_\soft \\
&= \int [\dt \CC^+][\dt \CC^-][\dt \CN^+][\dt \CN^-] \exp \left( - \frac{1}{2\pi e^2}\ln\frac{\L}{\mu}
		\int_{\mrr^2} \dt^2x \,\CN_a(x)\CN^a(x) \right) \\
&\quad \times \exp \left( - \frac{i}{e^2} \int_{\mrr^2} \dt^2 x
		 \left[ \left( \CC^{+a} (x) + \frac{1}{2} \CJ^{+a}(x) \right) \CN^+_a(x)
			- \left( \CC^{-a}(x) - \frac{1}{2} \CJ^{-a}(x) \right) \CN^-_a(x) \right] \right) .
\end{split}
\end{equation}
As before, we integrate out $\CN^\mathsf{avg}$ which yields a Dirac delta functional $\d ( \CC^+_a + \frac{1}{2}  \CJ^+_a - \CC^-_a + \frac{1}{2} \CJ^-_a )$ which implements the matching condition $\CC^+_a + \frac{1}{2}  \CJ^+_a = \CC^-_a - \frac{1}{2} \CJ^-_a$. We now integrate out the averaged edge mode $\CC_a$ which now yields a Dirac delta functional $\d(\CN_a )$ which sets $\CN_a=0$. The outcome of this procedure is then
\begin{equation}
    \begin{split}
        \dbra{0,+} \wt{\CO}_1\cdots \wt{\CO}_n \dket{0,-} |_\soft  = 1 . 
    \end{split}
\end{equation}
We immediately observe that the universal soft factor $\SS_0$ for dressed scattering amplitudes is entirely finite! All dependence on the infrared scale $\mu$ has completely disappeared. This shows that scattering amplitudes in FK dressed states are infrared finite!

\subsubsection{General Infrared Finite Dressings}

It is well known that the FK dressing \eqref{FK_dressing_gauge} is \emph{not} the only one that leads to infrared finite scattering amplitudes. We can also see this from our generalized SEA. To see this, consider a general dressing such that
\begin{equation}
\begin{split}
\label{Rk_gen}
\sum_{k=1}^n R^{\eta_k}_k &= - \frac{i}{e^2}  \int_{\mrr^2} \dt^2 x \left( C^{+a}(x) \CP_a^+(x) + C^{-a}(x) \CP_a^-(x) + N^{+a}(x) \CQ_a^+(x) + N^{-a}(x) \CQ_a^-(x) \right) . 
\end{split}
\end{equation}
We evaluate the path integral with this dressing. The integral over $\CN^\mathsf{avg}$ gives a Dirac delta functional that implements the matching condition
\begin{equation}
    \begin{split}
    \label{matching_cond_gen}
\CC^+_a(x) + \CQ^+_a (x) = \CC^-_a(x) - \CQ^-_a(x). 
    \end{split}
\end{equation}
Following this, the integral over $\CC^\mathsf{avg}$ gives a Dirac delta functional that implements the soft theorem in this dressed state
\begin{equation}
    \begin{split}
    \label{soft_thm_gen}
\CN_a(x)  = \CJ_a(x) - \CP^+_a(x) - \CP^-_a(x). 
    \end{split}
\end{equation}
Putting it altogether, the complete universal soft factor for the dressed amplitude is given by
\begin{equation}
\begin{split}
\label{soft_pi_gen_FK_11}
\dbra{0,+} \wt{\CO}_1\cdots \wt{\CO}_n \dket{0,-} |_\soft &= \exp \left( - \frac{1}{2\pi e^2}\ln\frac{\L}{\mu} \int_{\mrr^2} \dt^2x [   \CP^+_a + \CP^-_a - \CJ_a ]^2 \right) \\
&\qquad \times \exp \left( \frac{i}{e^2}\int_{\mrr^2} \dt^2 x \left( ( \CP^{+a} - \CJ^{+a} ) \CQ^+_a  - ( \CP^{-a} - \CJ^{-a} ) \CQ^-_a \right) \right)  .
\end{split}
\end{equation}
This dressed amplitude is infrared finite if and only if
\begin{equation}
\begin{split}
\label{Pa_constraint}
\CP^+_a(x) + \CP^-_a(x) = \CJ_a(x) . 
\end{split}
\end{equation}
This cancels the divergence proportional to $\ln \frac{\L}{\mu}$ and we are left with a finite soft factor. In particular, note that there is absolutely no constraint on $\CQ^\pm_a(x)$ as far as infrared finiteness is concerned. $\CQ^\pm_a(x)$ contributes to the matching condition \eqref{matching_cond_gen} and to the infrared finite part of the amplitude. We also note that for all such amplitudes, the leading soft-photon theorem is trivial since \eqref{soft_thm_gen} with \eqref{Pa_constraint} implies that $\CN_a(x) = 0$.

To summarize, we find that there are an infinite class of dressed states given by \eqref{Rk_gen} with \eqref{Pa_constraint} that all lead to infrared-finite scattering amplitudes.\footnote{We do not claim that \eqref{Rk_gen} is the most general dressing that leads to infrared finite amplitudes. One could perhaps include terms that are quadratic in the soft and edge modes. Another possible generalization includes dressing states with the radiative photon operators ${\hat \CO}_a(\o {\hat q}(x))$. These appear at $O(\ln \o)$ in \eqref{Oa_small_o}. These operators commute with the soft and Goldstone operators so they do not affect the soft factorization at all. We do not explore these generalizations here.} The FK dressing \eqref{FK_dressing_gauge} is a special case of this (with $\CP_a^\pm = \CJ_a^\pm$ and $\CQ_a^\pm = \pm \frac{1}{2} \CJ_a^\pm$).

\paragraph{Operator Interpretation of the Generalized Dressing:} It is instructive to explore the operator interpretation of \eqref{matching_cond_gen} and \eqref{Pa_constraint}.
\begin{itemize}[leftmargin=*]
    \item We start with \eqref{matching_cond_gen} and argue that despite appearances, it is in fact the \emph{same} as the operator matching condition \eqref{match_cond_gauge}. To see this, we note that the action of the \emph{out} part of \eqref{Rk_gen} on an \emph{out}-vacuum with eigenvalue $\CC^+$ is
\begin{equation}
    \begin{split}
\bra{\CC^+,+} \col \exp \left( - \frac{i}{e^2}  \int_{\mrr^2} \dt^2 x \left( C^{+a}(x) \CP_a^+(x) + N^{+a}(x) \CQ_a^+(x) \right) \right) \col . 
    \end{split}
\end{equation}
Here, $\col~\col$ denotes the soft-normal ordering discussed in Appendix \ref{app:Lorentz_inv_vac_state}. This moves the operator $C_a^+$ to the left and $N_a^+$ to the right. $C_a^+$ acts on the bra state and is simply replaced by $\CC_a^+$. On the other hand, from the property \eqref{commutators_gauge}, we find that the exponential operator associated $N_a^+$ acts on the bra state and shifts the \emph{out}-eigenvalue $\CC_a^+ \to \CC_a^+ + \CQ_a^+$. In the same way, the \emph{in} part of \eqref{Rk_gen} shifts the \emph{in}-eigenvalue $\CC_a^- \to \CC_a^- - \CQ_a^-$. The operator matching condition $C_a^+=C_a^-$ when evaluated in a correlator with \emph{in} and \emph{out} eigenvalues $\CC_a^- - \CQ_a^-$ and $\CC_a^+ + \CQ_a^+$ respectively, yields \eqref{matching_cond_gen}.

\item We next turn to \eqref{Pa_constraint} and show that this is exactly the result of \cite{Kapec:2017tkm}. To see this, we recall that the soft path integral evaluates the universal soft factor of the amplitude evaluated in an eigenstate of $N_a^\pm$ with eigenvalue $\CN_a^\pm=0$ (denoted by $\dket{0,\pm}$). The \emph{out} part of the dressing operator \eqref{Rk_gen} acting on the \emph{out}-vacuum state is
\begin{equation}
\begin{split}
\dbra{0,+} \col \exp \left( - \frac{i}{e^2}  \int_{\mrr^2} \dt^2 x \left( C^{+a}(x) \CP_a^+(x) + N^{+a}(x) \CQ_a^+(x) \right) \right) \col .
\end{split}
\end{equation}
It can be checked that the resultant state is also an eigenstate of $N_a^+$ with eigenvalue $\CN^\mathsf{out}_a(x) = \CP^+_a(x)$. Similarly, the \emph{in} part of \eqref{Rk_gen} acting on the \emph{in}-vacuum state is an eigenstate of $N_a^-$ with eigenvalue $\CN^\mathsf{in}_a(x) = - \CP^+_a(x)$. Using this, we can rewrite \eqref{Pa_constraint} as
\begin{equation}
\begin{split}
\CN^\mathsf{out}_a(x) - \CN^\mathsf{in}_a(x)  = \CJ_a(x) . 
\end{split}
\end{equation}
This is precisely equation (2.14) of \cite{Kapec:2017tkm}. It was argued in that paper that any amplitude satisfying the identity above is infrared finite. We have derived this result here using the generalized SEA.

\end{itemize}

\subsection{Gravity}
\label{sec:gen_sea_grav}

The generalized SEA in gravitational theories has exactly the same form as the one in gauge theory and it is given by
\begin{equation}
\begin{split}
\label{gen_sea_grav}
S_\eff[\CC^\pm,\CN^\pm] &= \frac{i}{2\pi \k^2} \ln\frac{\L}{\mu} \int_{\mrr^2} \dt^2x \,\CN_{ab}(x)\CN^{ab}(x) \\
&\qquad \qquad - \frac{1}{\k^2}\int_{\mrr^2} \dt^2x \left( \CC^{+ab}(x)\CN^+_{ab}(x) - \CC^{-ab}(x)\CN^-_{ab}(x) \right) , 
\end{split}
\end{equation}
where
\begin{equation}
\begin{split}
\CN_{ab}(x) \equiv \CN^+_{ab}(x) - \CN^-_{ab}(x) . 
\end{split}
\end{equation}
Unlike the gauge theory case, we do not yet have a complete first principles derivation of such an action, though steps in the right direction were taken in \cite{He:2024vlp}. The universal soft factor in a scattering amplitude is given by
\begin{equation}
\begin{split}
\label{soft_pi_gen_grav}
&\bra{\Psi_\mathsf{out},+} N_{a_1b_1} \cdots N_{a_mb_m} \CO_1\cdots\CO_n \ket{\Psi_\mathsf{in},-} |_\soft \\
&\qquad =\int [\dt \CC^+][\dt \CC^-][\dt \CN^+][\dt \CN^-] e^{iS_\eff[\CC^\pm,\CN^\pm]} 
\Psi_\mathsf{out}^*(\CC^+) \Psi_\mathsf{in}(\CC^-) \CN_{a_1b_1} \cdots \CN_{a_mb_m} \, \CU_1 \cdots \CU_n ,
\end{split}
\end{equation}
where the operators $\CU_k$ are given by
\begin{equation}
    \begin{split}
        \CU_1\cdots \CU_n &= \exp\left( \frac{i}{\k^2} \int_{\mrr^2} \dt^2x \left( \CC^{+ab}(x)\CJ^+_{ab}(x) + \CC^{-ab}(x)\CJ^-_{ab}(x) \right) \right)  .
    \end{split}
\end{equation}
Using this and \eqref{FK_dressing_grav}, the properties (1)--(4) can then easily be derived from this action in exactly the same way as we did for gauge theories.

\section{Summary and Outlook}
\label{sec:comments}

In this paper, we constructed a two-dimensional action that completely captures all universal infrared properties of scattering amplitudes in four-dimensional Abelian gauge and gravitational theories. The action constructed here is a generalization of the one introduced in \cite{Kapec:2021eug}. The SEA of \cite{Kapec:2021eug} can reproduce soft theorems and infrared divergences in Fock space scattering amplitudes, whereas the generalized SEA constructed here is able to reproduce soft theorems and infrared divergences in scattering amplitudes with \emph{arbitrary} \emph{in} and \emph{out} vacuum states. We used this to \emph{prove} that Faddeev-Kulish dressed states lead give rise to infrared finite scattering amplitudes, and were also able to construct more general infrared-finite dressing factors. Finally, our action also reproduces the antipodal matching conditions that are crucial to the scattering problem. This matching condition was input by hand in \cite{Kapec:2021eug}. The most remarkable aspect of these derivations is that they turn out to be remarkably straightforward, due to the Gaussian nature of the generalized SEA. The generalized SEA lives on the celestial sphere and is part of the full action (whatever that may be) of the celestial CFT that is dual to Abelian gauge and gravitational theories. Our results showcase the advantage that one gains from such a holographic perspective.

We end this section by briefly commenting on potential extensions of this work.

\paragraph{Higher Dimensions:} We focussed in this paper on the holographic structure of gauge and gravitational theories in four dimensions. The generalization to a higher dimensional action, however, is quite natural. The SEA for gauge theories in dimensions $D=d+2>4$ has already been constructed in \cite{Kapec:2021eug}. It is given by
\begin{equation}
\begin{split}
\label{sea_higherd}
S[\CC,\CN] = i a_d \int_{\mrr^d} \dt^d x \CN^a(x) \CN_a(x) + b_d \int_{\mrr^d} \dt^d x {\wt \CC}_a(x) \CN^a(x) , 
\end{split}
\end{equation}
where $a_d$ and $b_d$ are dimension-dependent constants given in \cite{Kapec:2021eug}, and $\CN_a(x) = \CN_a^+(x) - \CN_a^-(x)$. $a_d$ is cutoff dependent. ${\wt \CC}_a(x)$ is the shadow transform of $\CC_a(x)$ (see \cite{Kapec:2021eug} for a definition). As in four dimensions, $\CC_a(x)$ is the Goldstone mode for large gauge transformations and $\CN_a^\pm(x)$ are the asymptotic soft photon operators. From \eqref{sea_higherd}, we can immediately conjecture the \emph{generalized SEA} in higher dimensions to be
\begin{equation}
\begin{split}
S_\eff [\CC^\pm,\CN^\pm] = i a_d \int_{\mrr^d} \dt^d x \CN^a(x) \CN_a(x) + b_d \int_{\mrr^d} \dt^d x \left( {\wt \CC}^+_a(x) \CN^{+a}(x) - {\wt \CC}^-_a(x) \CN^{-a}(x)  \right) . 
\end{split}
\end{equation}
It is then easy to show that the entire calculation of Section \ref{sec:gen_sea} follows through trivially in higher dimensions as well. A similar generalization exists for gravity as well.

\paragraph{Infrared Divergent Phase:} The soft factor $\G_\gauge$ in \eqref{IR_factorization_gauge} also has an infrared divergent imaginary part that we have not presented in this paper. This imaginary part leads to an infrared divergent phase in the scattering amplitude. In their work \cite{Kulish:1970ut}, Faddeev and Kulish addressed this by constructing an operator $e^{ i \Upphi}$ that we use to dress asymptotic states (on top of the dressing $e^{R_k^\pm}$ in \eqref{FK_dressing_k_gauge}). The dressing $R_k^\pm$ cancels the real part of $\G_\gauge$ while the dressing $\Upphi$ cancels the imaginary part of $\G_\gauge$. However, unlike $R_k^\pm$ which dresses one-particle states (so each operator $\CO_k$ is dressed as $e^{R_k^\pm} \CO_k$ independent of all the other operators in the amplitude), $\Upphi$ dresses the entire multi-particle asymptotic state as a whole. Another way to state this is that we cannot write $\Upphi = \sum_k \Upphi_k$. Rather, the sum is over pairs of particles in the form $\Upphi = \sum_{k,k'} \Upphi_{k,k'}$. This more complicated ``pair-wise'' dressing that deals with the infrared divergent phase does not yet have a natural interpretation in celestial holography. We discuss this in detail in upcoming work \cite{Choi:2026xxx}.\footnote{The 2-point function of $C^\pm$ that follows from \eqref{gen_sea_gauge} is $\avg{ C^{\eta}(x) C^{\eta'}(x') } \sim \ln [ ( x - x' )^2 ]$. It was suggested in \cite{Gonzo:2022tjm} that the imaginary part of $\G_\gauge$ can be reproduced by modifying the two-point function to $\avg{ C^{\eta}(x) C^{\eta'}(x') } \sim \ln [ ( x - x' )^2 ] - i \pi \d_{\eta,\eta'}$. It turns out that this prescription works if and only if \emph{all} the particles in the scattering amplitude are massless (which is the setting of \cite{Gonzo:2022tjm}). When an amplitude involves two or more massive particles in the \emph{in} or \emph{out} state, this prescription does not correctly reproduce $\G_\gauge$.}

\paragraph{Subleading Soft Modes:} In this paper, we restricted our attention to the leading soft modes in gauge theory and gravity. It is now well known that similar soft modes appear at subleading orders in the soft expansion as well \cite{Low:1954kd,Low:1958sn,Gell-Mann:1954wra,Burnett:1967km,Cachazo:2014fwa,Lysov:2014csa,Kapec:2014opa,Hamada:2018vrw,Li:2018gnc,Sahoo:2018lxl,Strominger:2021mtt}. These are also associated to universal soft theorems and asymptotic symmetries. It should be possible to construct an SEA or a generalized SEA for these modes as well.

\paragraph{Connection to Schwinger-Keldysh:} By now, it is well established that infrared divergences in gauge and gravitational theories are associated to vacuum transitions in scattering amplitudes \cite{Kapec:2017tkm}. Quantum systems that undergo transitions are said to be ``out-of-equilibrium'', and one uses the Schwinger-Keldysh (SK) formalism to describe them. Correlators in such a system is described by the SK action that takes the form
\begin{equation}
\begin{split}
\label{SK_action}
S_{\mathsf{SK}} [q_+,q_-] = I_\mathsf{FV}[q_+,q_-] + S[q_+] - S[q_-]. 
\end{split}
\end{equation}
Here, $S[q]$ is the action of the system.\footnote{In the SK formalism, one typically computes \emph{in}-\emph{in} or out-of-time-ordered correlators. In the SK action, we double the fields in the system $q \to q_\pm$. The action $S[q_+]$ performs forward time evolution and $S[q_-]$ performs the backward time-evolution.} The Feynman-Vernon influence function $I_\mathsf{FV}[q_+,q_-]$ describes the interaction of the open system with the environment. The SK action above is remarkably similar in structure to the generalized SEA \eqref{gen_sea_gauge} if we identify
\begin{equation}
\begin{split}
S[q] = - \frac{1}{e^2} \int_{\mrr^2} \dt^2 x \CC^a \CN_a , \qquad I_\mathsf{FV}[q_+,q_-] = \frac{i}{2\pi e^2}\ln\frac{\L}{\mu} \int_{\mrr^2} \dt^2 x \,(\CN^+_a - \CN^-_a )^2 .
\end{split}
\end{equation}
It would be interesting to explore the extent to which this similarity between actions can be taken seriously.

\section*{Acknowledgements}

We would like to thank Laurent Freidel, Kelian H\"aring, Temple He, Richard M.\ Myers, Andrea Puhm, Kamran Salehi Vaziri and Kathryn Zurek for useful discussions. SC and PM are supported by the European Research Council (ERC) under the European Union’s Horizon 2020 research and innovation programme (grant agreement No 852386). SC and PM are supported by the European Research Council under the European Union’s Seventh Framework Programme (FP7/2007-2013),
ERC Grant agreement ADG 834878. This work was supported by the Simons Collaboration on Celestial Holography.

\appendix

\section{Vacuum Hilbert Space in Gauge Theories}
\label{app:cps_gauge}

\subsection{Soft Commutators}
\label{app:soft_comm}

In this Appendix, we derive the soft commutators \eqref{commutators_gauge}. To do this, we use the covariant phase space formalism \cite{Crnkovic:1986ex, Lee:1990nz, Wald:1993nt, Iyer:1994ys, Barnich:2001jy, Harlow:2019yfa, He:2020ifr, Shi:2020csw}. Here, we start with the action of the gauge theory which has the form
\begin{equation}
\begin{split}
\label{app:gauge_action}
S = - \frac{1}{4e^2} \int_{M_4} \dt^4 X F^{\mu\nu} F_{\mu\nu} + S_\mat . 
\end{split}
\end{equation}
where the matter action $S_\mat$ includes the kinetic terms for the matter fields and all the interaction in the theory. The current $J_\mu(X)$ appearing in \eqref{Maxwell_Eq} in given by
\begin{equation}
\begin{split}
\label{app:current_def}
J_\mu(X) \equiv - \frac{\p S_\mat}{\p A^\mu(X)}  . 
\end{split}
\end{equation}
This is conserved on-shell due to gauge invariance of the matter action. We vary the action, extract the boundary term, vary a second time, and integrate it over a Cauchy slice $\S$ to obtain the symplectic form. These are
\begin{equation}
\begin{split}
\label{app:sympform_gauge}
\O = - \frac{1}{e^2} \int_{\S} \dt \S \, n^\mu \d F_{\mu\nu} \w \d A^\nu + \O_\S^\mat .
\end{split}
\end{equation}
On the asymptotic Cauchy slices $\S^\pm = \CI^\pm \cup i^\pm$ the gauge and matter contributions to the symplectic potentials separate out. Consequently, to construct the asymptotic Hilbert space, we can focus on each term separately. Moving to flat null coordinates \eqref{flat_null_coord} and evaluating the symplectic potential on $\S^\pm$, we find
\begin{equation}
\begin{split}
\label{app:Omega_ip_gauge}
\O^\gauge_{\S^\pm} = \frac{1}{e^2} \int_{\CI^\pm} \dt u \dt^2 x  \p_u \d A_a^\pm(u,x) \w \d A^{\pm a}(u,x)  , \qquad A^\pm_a(u,x) \equiv \lim_{r \to \pm \infty} A_a(u,x,r) . 
\end{split}
\end{equation}
We now rewrite this using the mode expansion \eqref{mode_exp_gauge}. The large $|r|$ limit of the mode expansion can be evaluated using the stationary phase approximation, and we find
\begin{equation}
\begin{split}
\label{app:mode_exp_gauge}
A^\pm_a(u,x) = - \frac{ie}{8\pi^2} \int_0^\infty \dt \o \big[ \CO_a^\pm (\o,x) e^{ - i \o u } - \CO_a^{\pm\dagger}(\o,x) e^{  i \o u } \big] .
\end{split}
\end{equation}
Substituting \eqref{Oa_small_o} into \eqref{app:mode_exp_gauge}, we find that $A_a^\pm(u,x)$ has the form
\begin{equation}
\begin{split}
\label{app:mode_exp_gauge_1}
A^\pm_a(u,x) = C_a^\pm(x) + {\hat A}_a^\pm(u,x) , \qquad {\hat A}_a^\pm(\mp\infty,x) = 0  , \qquad {\hat A}_a^\pm(\pm\infty,x) = \pm N_a^\pm(x).
\end{split}
\end{equation}
Substituting this into \eqref{app:Omega_ip_gauge} and extracting the soft contribution, we find
\begin{equation}
\begin{split}
\O^{\soft}_{\S^\pm} = \frac{1}{e^2} \int_{\mrr^2} \dt^2 x  \d N_a^\pm(x) \w \d C^{\pm a}(x) = - \frac{1}{e^2} \int_{\mrr^2} \dt^2 x  \d N^\pm(x) \w \p^2 \d C^\pm(x) . 
\end{split}
\end{equation}
where in the last equality, we used \eqref{no_magnetic_charges}. Inverting this, we find the quantum commutators \eqref{commutators_gauge}.

\subsection{Lorentz-Invariant Vacuum State}
\label{app:Lorentz_inv_vac_state}

Using the soft symplectic form, we can determine the soft contribution to the Noether charges that generate Poincar\'e transformations. To do this, we note that the vector fields that generate translations and Lorentz transformations are
\begin{equation}
\begin{split}
\xi_f &\equiv - \chi^\mu \p_\mu = f \p_u + \frac{1}{2} \p^2 f \p_r - \frac{1}{r} \p^a f \p_a , \\
\zeta_Y &\equiv - \o^\mu{}_\nu X^\nu \p_\mu =  \frac{1}{2} ( \p \cdot Y ) ( u \p_u - r \p_r ) + \left( Y^a - \frac{u}{2r} \p^a ( \p \cdot Y ) \right) \p_a,
\end{split}
\end{equation}
where
\begin{equation}
\begin{split}
f(x) = \chi^\mu {\hat q}_\mu (x) , \qquad Y^a(x) = \o^{\mu\nu} {\hat q}_\mu(x) \p^a {\hat q}_\nu (x) . 
\end{split}
\end{equation}
$Y^a(x)$ is a global conformal Killing vector on $S^2$. Under Poincar\'e transformations, the gauge field transforms via a Lie derivative, $\d_f A_\mu = \CL_{\xi_f} A_\mu$, $\d_Y A_\mu = \CL_{\zeta_Y} A_\mu$. Moving to flat null coordinates and taking the limit to $\CI^\pm$, we find
\begin{equation}
\begin{split}
\d_f A^\pm_a = f \p_u A_a^\pm , \qquad \d_Y A^\pm_a = \left( \CL_Y + \frac{1}{2} ( \p \cdot Y ) u \p_u \right) A^\pm_a . 
\end{split}
\end{equation}
Finally, using \eqref{app:mode_exp_gauge_1}, we find
\begin{equation}
\begin{split}
( \d_f N^\pm_a , \d_f C^\pm_a ) = 0 , \qquad ( \d_Y N^\pm_a , \d_Y C^\pm_a ) = ( \CL_Y N^\pm_a , \CL_Y C^\pm_a ) . 
\end{split}
\end{equation}
The first equality implies that the soft operators are invariant under time and space translations. Consequently, the operators commute with the Hamiltonian $H$ of the gauge theory, and therefore, span the vacuum Hilbert space. From the second property, we can determine the soft contribution to the Lorentz charge, which satisfies $\d J^{\pm,\soft}_Y = \CI_{\d_Y} \O^{\soft}_{\S^\pm}$. It follows from this that
\begin{equation}
\begin{split}
\label{Lorentz_charge}
J^{\pm,\soft}_Y = \frac{1}{e^2} \int_{\mrr^2} \dt^2 x \CL_Y C^\pm_a(x) N^{\pm a}(x) .
\end{split}
\end{equation}
Note that in the soft Lorentz charge, we have ordered the operators so that $N$ is on the right and $C$ is on the left. We refer to this ordering as the \emph{soft normal-ordering}. All operators products are normal-ordered in this way. From \eqref{Lorentz_charge}, it follows immediately that the eigenstate of $N^\pm(x)$ with eigenvalue $\p_a \CN(x) = 0$ is Lorentz-invariant.\footnote{In previous literature, \cite{He:2020ifr} the authors used an opposite normal-ordering in which all the $C$'s are on the right and $N$'s are on the left. With this choice, the Lorentz-invariant states are eigenstates of $C^\pm$ with eigenvalue $\p_a \CC(x) = 0$.}

\subsection{LSZ Reduction Formula \texorpdfstring{$\to$}{to} Boundary Operators}
\label{app:LSZ_bdy}

In this section, we prove \eqref{soft_limit_op_gauge}. We start with the LSZ reduction formula \eqref{LSZ_def_gauge} for the photon
\begin{equation}
\begin{split}
\label{app:LSZ_gauge}
\CO_a(q) = - \frac{i}{e} \ve^\mu_a(q) \int_{M_4} \dt^4 X e^{ - i q \cdot X } \p^2 A_\mu (X).
\end{split}
\end{equation}
To simplify this, we use the mode expansion \eqref{mode_exp_gauge}, which we recall here
\begin{equation}
\begin{split}
\label{app:mode_exp}
A_\mu(X) = e \int_{\mrr^3} \frac{\dt^3q}{(2\pi)^3} \frac{1}{2q^0} \big[ a_\mu (t,q) e^{ i q \cdot X} + a_\mu^\dagger(t,q) e^{ - i q \cdot X} \big] .
\end{split}
\end{equation}
It follows that 
\begin{equation}
\begin{split}
\CO_a(q) &= \frac{\ve^\mu_a(q)}{2|\vec{q}\,|} \big[ ( q^0 + |\vec{q}\,| ) a_\mu (t,\vec{q}\,) e^{ i ( q^0 - |\vec{q}\,| ) t } + ( q^0 - |\vec{q}\,| ) a_\mu^\dagger(t,-\vec{q}\,) e^{ i ( q^0 + |\vec{q}\,| ) t }  \big] \bigg|_{t=-\infty}^{t=+\infty}
\end{split}
\end{equation}
where we used the following property to simplify our result
\begin{equation}
\begin{split}
\lim_{t \to \pm \infty} \p_t  a_\mu (t,\vec{q}\,)  = 0 .
\end{split}
\end{equation}
Using \eqref{a_large_t_gauge}, we can take the large $t$ limit and find
\begin{equation}
\begin{split}
\CO_a(q) = \t(q^0) [ \CO_a^+(q) - \CO_a^-(q) ] + \t(-q^0) [ \CO_a^{-\dagger}(-q) - \CO_a^{+\dagger}(-q) ] ,
\end{split}
\end{equation}
where $\t(x)$ is the Heaviside theta function. We now parameterize $q^\mu = \o {\hat q}^\mu(x)$ and take the soft limit $\o \to 0$ (after multiplying by $\o$). The soft limit is then simplify using \eqref{Oa_small_o}. Hermiticity of $N^\pm$ and $C^\pm$ implies that the limits $\o \to 0^+$ and $\o \to 0^-$ are both equal and we find
\begin{equation}
\begin{split}
\lim_{\o \to 0} \left[ \o \CO_a(\o {\hat q}(x)) \right] = - \frac{4\pi}{e} [ N_a^+(x) - N_a^-(x) ] . 
\end{split}
\end{equation}
This is precisely \eqref{soft_limit_op_gauge}.

\subsection{Matching Conditions}
\label{app:matching_cond}

In this section, we derive the matching conditions \eqref{match_cond_gauge}. This was accomplished by Campiglia and Eyheralde in \cite{Campiglia:2017mua} and we reproduce their result here, though approach differs slightly (see also \cite{Capone:2022gme, Compere:2025tzr}). To emphasize the fact these matching conditions are in fact \emph{antipodal} in nature, we will work in spherical coordinates instead of flat null coordinates. Define
\begin{equation}
\begin{split}
t = \rho \sinh \tau , \qquad \vec{x}  = \rho \cosh \tau {\hat x} , \qquad \tau \in \mrr , \qquad \rho > 0  , \qquad |{\hat x}| = 1 .
\end{split}
\end{equation}
These coordinates cover the region of Minkowski spacetime outside the lightcone centered at $X^\mu=0$ (defined by $|\vec{x}|>|t|$). The Minkowski metric is
\begin{equation}
\begin{split}
\label{app:dS_metric}
\dt s^2 = \dt \rho^2 + \rho^2 \dt s_{\dS}^2 , \qquad \dt s_{\dS}^2 =  - \dt \tau^2  + \cosh^2 \tau \dt \O_2^2.
\end{split}
\end{equation}
$\dt \O_2^2 = \dt {\hat x} \cdot \dt {\hat x}$ is the unit sphere metric. The induced metric on a constant $\rho$ slice is the de Sitter metric (with $L_\dS = \rho$). Spatial infinity $i^0$ is reached by taking $\rho \to \infty$. $\rho = 0$ is the light cone $X^\mu X_\mu = 0$. On each de Sitter slice, the codimension two boundaries at $\tau \to \pm \infty$ are $\p \S^\pm$.

To derive the matching conditions, we consider solutions to Maxwell's equations near spatial infinity. Point charges reach null infinity (if they are massless) or timelike infinity (if they are massive), but not spatial infinity. Consequently, near $\rho=\infty$, we can assume $J_\mu = 0$. We work in radial gauge ($A_\rho = 0$) and decompose the angular components of the gauge field into two scalars as
\begin{equation}
\begin{split}
\label{app:AA_decomp}
A_A(X) = D_A \phi(X) + \e_{AB} D^B \psi(X) . 
\end{split}
\end{equation}
We expand the remaining components in a basis of spherical harmonics,
\begin{equation}
\begin{split}
A_\tau (\rho,\tau,{\hat x}) &= \sum_{\ell=0}^\infty \sum_{m=-\ell}^{\ell} A_{\tau;\ell,m}(\rho,\tau) Y_{\ell,m}({\hat x}), \\
\phi (\rho,\tau,{\hat x}) &= \sum_{\ell=1}^\infty \sum_{m=-\ell}^{\ell} \phi_{\ell,m}(\rho,\tau) Y_{\ell,m}({\hat x}) , \\
\psi (\rho,\tau,{\hat x}) &= \sum_{\ell=1}^\infty \sum_{m=-\ell}^{\ell} \psi_{\ell,m}(\rho,\tau) Y_{\ell,m}({\hat x}).
\end{split}
\end{equation}
Since only derivatives of $\phi$ and $\psi$ appear in \eqref{app:AA_decomp}, the spherical harmonic expansion for these fields include only the $\ell \geq 1$ modes.

Next, we expand the fields at large $\rho$, 
\begin{equation}
\begin{split}
A_{\tau;\ell,m} (\rho,\tau) &= A_{\tau;\ell,m}^\0(\tau) + \rho^{-1} A_{\tau;\ell,m}^\1(\tau) + O(\rho^{-2}) , \\
\phi_{\ell,m}(\rho,\tau) &= \phi_{\ell,m}^\0(\tau)  + O(\rho^{-1}) , \qquad \psi_{\ell,m}(\rho,\tau) = \psi_{\ell,m}^\0(\tau) + O(\rho^{-1}) .
\end{split}
\end{equation}
We plug in this expansion into Maxwell's equations and extract the leading order differential equations. The smooth solutions are
\begin{equation}
\begin{split}
\label{app:Maxwell_sol}
A_{\tau;\ell,m}^\0(\tau) = \p_\tau \phi_{\ell,m}^\0(\tau) , \qquad A^\1_{\tau;\ell,m} &= a_{\ell,m}^\1 \frac{P_\ell^1 ( \tanh \tau )}{\cosh^2 \tau} , \qquad \psi^\0_{\ell,m} (\tau) = b^\0_{\ell,m} P_\ell(\tanh \tau) . 
\end{split}
\end{equation}
where $P_\ell^m(t)$ is the associated Legendre polynomial. $\phi_{\ell,m}^\0(\tau)$ is unfixed by Maxwell's equations. We require that it satisfies, 
\begin{equation}
\begin{split}
\phi_{\ell,m}^\0(+\infty) = (-1)^\ell \phi_{\ell,m}^\0(-\infty)  .
\end{split}
\end{equation}
From \eqref{app:Maxwell_sol}, we find that $\psi^\0$ satisfies the same property
\begin{equation}
\begin{split}
\psi_{\ell,m}^\0(+\infty) = (-1)^\ell \psi_{\ell,m}^\0(-\infty) .
\end{split}
\end{equation}
The explicit factor of $(-1)^\ell$ in the properties above imply that the spacetime fields are antipodal matched! To be precise, the property $Y_{\ell,m}(-{\hat x}) = (-1)^\ell Y_{\ell,m}({\hat x})$ implies
\begin{equation}
\begin{split}
\phi^\0 (+\infty,{\hat x}) = \phi^\0 (-\infty,-{\hat x}) , \qquad  \psi^\0 (+\infty,{\hat x}) = \psi^\0 (-\infty,-{\hat x}) . 
\end{split}
\end{equation}
We note again that the antipodal matching condition on $\phi$ is imposed by hand, but the one on $\psi$ follows from Maxwell's equations. Maxwell's equations also implies a second matching condition
\begin{equation}
\begin{split}
\label{second_match_cond}
\lim_{\tau \to \infty} \left( \sinh^3 \tau A^\1_{\tau;\ell,m} \right)  = (-1)^\ell \lim_{\tau \to - \infty} \left( \sinh^3 \tau  A^\1_{\tau;\ell,m} \right)  . 
\end{split}
\end{equation}
We now use the following result derived in\cite{Campiglia:2017mua} (Eqs. (3.7) and (3.8))\footnote{Note that $\tau_\mathsf{there} = \sinh ( \tau_\mathsf{here} )$.}
\begin{equation}
\begin{split}
( r^2 F_{ru} )|_{\p \S^+}  = \lim_{\tau \to + \infty} \lim_{\rho \to \infty} \left( \sinh^3 \tau \rho  F_{\rho \tau} \right) , \qquad ( r^2 F_{rv} )|_{\p \S^-}  = \lim_{\tau \to - \infty} \lim_{\rho \to \infty}  \left( \sinh^3 \tau  \rho F_{\rho \tau} \right) . 
\end{split}
\end{equation}
This relates the boundary field strength in our de Sitter coordinates to the boundary field strength in advanced and retarded coordinates ($u=t-r$, $v=t+r$, and $r=|\vec{x}|$). Using this and \eqref{second_match_cond}, we find
\begin{equation}
\begin{split}
( r^2 F_{ru} )|_{\p \S^+} ({\hat x}) = ( r^2 F_{rv} )|_{\p \S^-} (-{\hat x}).
\end{split}
\end{equation}
After moving back to flat null coordinates, it can be shown that this is precisely the second matching condition in \eqref{match_cond_gauge} \cite{He:2019jjk}.

\bibliographystyle{utphys}
\bibliography{FK-bib}

\end{document}